\documentclass[aps,pra,reprint,amsmath,amssymb,groupedaddress,showpacs]{revtex4-1}
\usepackage{graphicx}
\usepackage{float}
\usepackage{dcolumn} 
\usepackage{bm}      
\usepackage[bookmarks=false]{hyperref} 
\usepackage[font={small}]{caption} 
\hypersetup{
    unicode=false,          
    pdftoolbar=true,        
    pdfmenubar=true,        
    pdffitwindow=true,      
    pdfstartview={},        
    pdftitle={Optomechanical Entanglement of Remote Microwave Cavities}, 
    pdfauthor={Samuel R. Hedemann and B. D. Clader},              
    pdfsubject={Entanglement in Quantum Optics}, 
    pdfcreator={},          
    pdfproducer={},         
    pdfkeywords={Optomechanical,} {Entanglement,} {Cavities,} {Microwave,} {Logarithmic Negativity,} {Ent,} {Superconducting Qubits,} {Quantum Optics}, 
    pdfnewwindow=true,      
    colorlinks=true,        
    linkcolor=black,        
    citecolor=black,        
    filecolor=black,        
    urlcolor=black          
}
\newcommand{\Sec}[1]{\hyperref[sec:#1]{Sec.{\kern 2pt}\ref*{sec:#1}}}
\newcommand{\Section}[1]{\hyperref[sec:#1]{Section~\ref*{sec:#1}}}
\newcommand{\Fig}[2][]{\hyperref[fig:#2]{Fig.{\kern 2pt}\ref*{fig:#2}#1}}
\newcommand{\Figure}[2][]{\hyperref[fig:#2]{Figure~\ref*{fig:#2}#1}}
\newcommand{\App}[1]{\hyperref[sec:#1]{App.{\kern 2pt}\ref*{sec:#1}}}
\newcommand{\Appendix}[1]{\hyperref[sec:#1]{Appendix~\ref*{sec:#1}}}
\newcommand{\Eq}[1]{\hyperref[eq:#1]{(\ref*{eq:#1})}}
\newcommand{\Table}[2][]{\hyperref[tab:#2]{Table~\ref*{tab:#2}#1}}
\begin{document}
\title{Optomechanical Entanglement of Remote Microwave Cavities}
\author{Samuel R. Hedemann}
\email[]{samuel.hedemann@gmail.com}
\author{B. D. Clader}
\email[]{dave.clader@jhuapl.edu}
\affiliation{The Johns Hopkins University Applied Physics Laboratory, Laurel, MD 20723, USA}
\date{\today}
\begin{abstract}
We examine the entanglement properties of a system that represents two driven microwave cavities each optomechanically coupled to two separate driven optical cavities which are connected by a single-mode optical fiber. The results suggest that it may be possible to achieve near-maximal entanglement of the microwave cavities, thus allowing a teleportation scheme to enable interactions for hybrid quantum computing of superconducting qubits with optical interconnects.
\end{abstract}
\maketitle
\section{\label{sec:I}Introduction}
Classical long-haul information networks utilize optical fibers for their low-loss and low-noise characteristics. Similarly, in the quantum regime, photons in optical fibers provide a proven and robust method to transmit quantum information reliably over long distances and at room temperature. Superconducting qubits, a leading candidate for a future quantum computer, operate in the microwave regime and must be cooled to milli-Kelvin temperatures for quantum operation. Off-chip connectivity at room temperature would be highly desirable in order to optically link superconducting chips. However, this requires a high-fidelity microwave-to-optical transducer that can be fiber-coupled into a quantum network. 

The fundamental problem is then to convert a microwave qubit to an optical qubit, which was investigated via optomechanical resonators by \cite[]{TiWa,SRSL,LTi1,BAMT,Clad}, and found to be feasible, while the concept of transferring states through optical connections in the presence of decoherence has been around for a while \cite[]{Pell,CZKM,EKCZ,SeMB,ReRe,KMWR,ABPZ}.

The next stage of the problem is to couple two such optomechanical resonators by an optical fiber, as treated in \cite[]{Clad} which showed that high-fidelity adiabatic state transfer is possible under realistic conditions, as did \cite{YYSD}.

A logical next step is to ask, \textit{can we achieve strong entanglement between optically networked microwave cavities, to use as a resource for quantum teleportation to improve state-transfer fidelity?} Thus, the goal of the present work is to answer this question.  While \cite[]{LinT} investigated the possibility of entangling two cavities connected by an optomechanical resonator, here we seek to entangle the two microwave cavities at either end of two different optomechanical resonators connected by an optical fiber, as seen in \Fig{1} in \Sec{II}.
\section{\label{sec:II}Model}
\begin{figure}[H]
\centering
\includegraphics[width=1.00\linewidth]{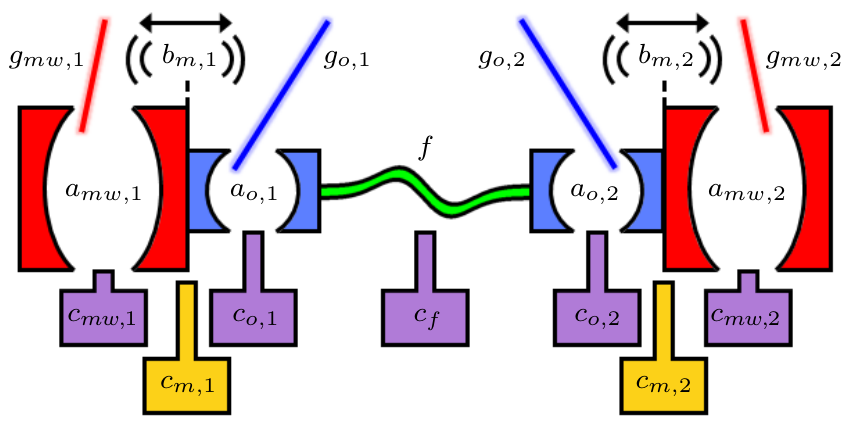}%
\vspace{-7pt}
\caption[]{(color online) Two optomechanical oscillators each consisting of two laser-driven cavities coupled by a two-sided oscillating mirror, both connected by an optical fiber, and each primary subsystem coupled to its own isolated bath. Mode operators are shown near each component, and $g_k$ are driving laser strengths.}
\label{fig:1}
\end{figure}

The overall model is 14 subsystems (modes) consisting of seven \textit{primary} modes each individually coupled to one of seven noninteracting baths which are the \textit{secondary} modes.  \Figure{1} depicts the total system.
The linearized interaction-picture Hamiltonian is
\begin{equation}
H = H_{om,1}  + H_f  + H_{om,2}  + H_d ,
\label{eq:1}
\end{equation}
where the time-dependence is suppressed, and where
\begin{equation}
\begin{array}{*{20}l}
   {H_{om,1} } &\!\! { = \hbar \left( \begin{array}{l}
 g_{mw,1} (a_{mw,1} ^{\dag}  b_{m,1}  + a_{mw,1} b_{m,1} ^{\dag}  ) \\ 
  + g_{o,1} (a_{o,1} ^{\dag}  b_{m,1}  + a_{o,1} b_{m,1} ^{\dag}  ) \\ 
 \end{array} \right)}  \\
   {H_{om,2} } &\!\! { = \hbar \left( \begin{array}{l}
 g_{o,2} (a_{o,2} ^{\dag}  b_{m,2}  + a_{o,2} b_{m,2} ^{\dag}  ) \\ 
  + g_{mw,2} (a_{mw,2} b_{m,2}  + a_{mw,2} ^{\dag}  b_{m,2} ^{\dag}  ) \\ 
 \end{array} \right)\!,}  \\
\end{array}
\label{eq:2}
\end{equation}
are the optomechanical terms where the beam-splitter terms of the form $\hbar g(a^{\dag}  b + ab^{\dag}  )$ correspond to cavities driven in the ``low sideband,'' meaning that the driving lasers are detuned below the cavity resonance frequency with detuning $\Delta  \equiv \omega _L  - \omega _c  \approx  - \omega _m $ where $\omega _L$ is the laser field's angular frequency, $\omega _c $ is the resonsant angular frequency of the cavity being driven, and $\omega _m$ is the resonant angular frequency of the mechanical oscillator of the local optomechanical system (two cavities coupled by a moveable two-sided mirror).  The term with form $\hbar g(ab + a^{\dag}  b^{\dag}  )$ is a two-mode squeezing Hamiltonian corresponding to driving in the ``high sideband,'' meaning that the second microwave cavity's driving laser is detuned above the cavity resonance frequency as $\Delta  \equiv \omega _L  - \omega _c  \approx  + \omega _m$.  Thus we call this the ``LLLH model'' since three cavities are driven low, and the one at the end is driven high.

The reason for driving one of the four cavities in the high sideband is that if our goal is to generate two-mode squeezing between the two microwave cavities, then we want to achieve the \textit{Bogoliubov modes}, which are \cite[]{LinT}
\begin{equation}
\begin{array}{*{20}l}
   {\beta _1 } &\!\! { = \cosh (r)a_{mw,1}  + i\sinh (r)a_{mw,2} ^{\dag}  }  \\
   {\beta _2 ^{\dag}  } &\!\! { =  - i\sinh (r)a_{mw,1}  + \cosh (r)a_{mw,2} ^{\dag},  }  \\
\end{array}
\label{eq:3}
\end{equation}
which lead to two-mode squeezing on the vacuum, where $r\in[0,\infty)$ is the \textit{squeezing parameter}.  The fact that the Bogoliubov modes are combinations of annihilation and creation operators means that the cavities being squeezed must have opposite sideband driving, and it was found that such modes only emerged when the other two cavities were driven in the low sideband.

The single-mode optical fiber's lone Hamiltonian is
\begin{equation}
H_f  = \hbar g_f ((a_{o,1}  - a_{o,2} )f^{\dag}   + (a_{o,1} ^{\dag}   - a_{o,2} ^{\dag}  )f),
\label{eq:4}
\end{equation}
and the dissipation term is
\begin{equation}
\begin{array}{*{20}l}
   {H_d =} &\!\! { H_{do,1}  + H_{dm,1}  + H_{dmw,1}  + H_{d{\kern 0.1pt}f} }  \\
   {} &\!\! { + H_{do,2}  + H_{dm,2}  + H_{dmw,2} ,}  \\
\end{array}
\label{eq:5}
\end{equation}
where each of these terms has the form
\begin{equation}
\begin{array}{*{20}l}
   {H_{dk} (t) = } &\!\! {\hbar \int_{ - \infty }^\infty  { (\omega  - \omega _k )c_k ^{\dag}  (\omega ,t)c_k (\omega ,t)d\omega } } \\
   {} &\!\! { + i\hbar \int_{ - \infty }^\infty  { g_{dk} (a_k ^{\dag}  (t)c_k (\omega ,t) - a_k (t)c_k ^{\dag}  (\omega ,t))d\omega }},  \\
\end{array}
\label{eq:6}
\end{equation}
where the fact that coupling strengths $g_{dk}$ are shown as frequency-independent is the ``first Markov approximation,'' and this form is the result of the rotating wave approximation (RWA).  Note that in \Eq{6}, $a$ is an abbreviation for \textit{all} main-system annihilation (mode) operators, and $k$ is a general index that can take on any label of each subsystem, and where the ordering of these labels will be defined in \Sec{II.A}.  Also note that \cite[]{GaCo} specifies the opposite order of terms in the second large term of \Eq{6}, likely because the naming convention in \cite[]{LinT} caused an exchange of terms, leading to an apparently unimportant sign-flip in the coupling strengths of the baths.  We leave it here, as in \cite[]{Clad}, for consistency with \cite[]{LinT}.
\subsection{\label{sec:II.A}Langevin Equation and Solution}
The Langevin equation for this system is
\begin{equation}
i{\textstyle{{d\mathbf{v}(t)} \over {dt}}} = M(t)\mathbf{v}(t) + i\sqrt{K} \mathbf{v}_{\text{in}} (t),
\label{eq:7}
\end{equation}
as derived in \App{App.A}, where, suppressing time,
\begin{equation}
{\!\!}\begin{array}{*{20}l}
   {\mathbf{v}} &\!\!\! { \equiv\! (a_{o,1} ,b_{m,1} ,a_{mw,1} ,f,a_{o,2} ,b_{m,2}, a_{mw,2} ^{\dag}  )^T }  \\
   {\mathbf{v}_{\text{in}}} &\!\!\! { \equiv\! (a_{o,1,\text{in}} ,b_{m,1,\text{in}} ,a_{mw,1,\text{in}} ,f,a_{o,2,\text{in}} ,b_{m,2,\text{in}}, a_{mw,2,\text{in}} ^{\dag}  )^{T}\!\!\!\!, }  \\
\end{array}
\label{eq:8}
\end{equation}
and the diagonal damping matrix is
\begin{equation}
K \equiv {\textstyle{1 \over {2\pi }}}\text{diag}(\begin{array}{*{20}c}
   {\kappa _{o,1} } & {\kappa _{m,1} } & {\kappa _{mw,1} } & {\kappa _f } & {\kappa _{o,2} } & {\kappa _{m,2} } & {\kappa _{mw,2} }  \\
\end{array}),
\label{eq:9}
\end{equation}
and the time-dependent dynamics matrix is
\begin{equation}
M{\kern -1.5pt}\equiv{\kern -4.0pt}\left(\!\! {\begin{array}{*{20}c}
   {  {\textstyle{{-i\kappa _{o,1} } \over 2}}} &\!\!\! {g_{o,1} } &\!\!\!  \cdot  &\!\!\! {g_f } &\!\!\!  \cdot  &\!\!\!  \cdot  &\!\!\!  \cdot   \\
   {g_{o,1} } &\!\!\! {  {\textstyle{{-i\kappa _{m,1} } \over 2}}} &\!\!\! {g_{mw,1} } &\!\!\!  \cdot  &\!\!\!  \cdot  &\!\!\!  \cdot  &\!\!\!  \cdot   \\
    \cdot  &\!\!\! {g_{mw,1} } &\!\!\! {  {\textstyle{{-i\kappa _{mw,1} } \over 2}}} &\!\!\!  \cdot  &\!\!\!  \cdot  &\!\!\!  \cdot  &\!\!\!  \cdot   \\
   {g_f } &\!\!\!  \cdot  &\!\!\!  \cdot  &\!\!\! {  {\textstyle{{-i\kappa _f } \over 2}}} &\!\!\! { - g_f } &\!\!\!  \cdot  &\!\!\!  \cdot   \\
    \cdot  &\!\!\!  \cdot  &\!\!\!  \cdot  &\!\!\! { - g_f } &\!\!\! {  {\textstyle{{-i\kappa _{o,2} } \over 2}}} &\!\!\! {g_{o,2} } &\!\!\!  \cdot   \\
    \cdot  &\!\!\!  \cdot  &\!\!\!  \cdot  &\!\!\!  \cdot  &\!\!\! {g_{o,2} } &\!\!\! {  {\textstyle{{-i\kappa _{m,2} } \over 2}}} &\!\!\! {g_{mw,2} }  \\
    \cdot  &\!\!\!  \cdot  &\!\!\!  \cdot  &\!\!\!  \cdot  &\!\!\!  \cdot  &\!\!\! { - g_{mw,2} } &\!\!\! {  {\textstyle{{-i\kappa _{mw,2} } \over 2}}}  \\
\end{array}}\!\! \right){\kern -2.5pt},
\label{eq:10}
\end{equation}
where the dots represent zeros, and we suppressed the time-dependence in $M$ and the driving-laser strengths \smash{$g_{mw,1}$}, \smash{$g_{mw,2}$}, \smash{$g_{o,1}$}, \smash{$g_{o,2}$}, and the damping strengths are \smash{$\kappa _k  \equiv 2\pi g_{dk} ^2$}. The order of the subsystems was chosen to give $M$ as close to a block-diagonal form as possible.

The solution to the Langevin equation in \Eq{7} is
\begin{equation}
\mathbf{v}(t) = \tau (t,0)\mathbf{v}(0) + \int_0^t {\tau (t,t')\sqrt{K} \mathbf{v}_{\text{in}} (t')} dt',
\label{eq:11}
\end{equation}
where $\tau (t,t')$ is the propagator
\begin{equation}
\tau (t,t') \equiv \tau \left\{ {e^{ - i\int_{t'}^t {M(t'')dt''} } } \right\},
\label{eq:12}
\end{equation}
where $\tau\{\cdot\}$ is the time-ordering operator.
\subsection{\label{sec:II.B}Determination of Parameter Values and Pre-Optimization}
There are several preliminary steps to take before looking at entanglement.  First, we have to find a set of physically realistic starting values for the parameters of the problem.  Then we have to look for nearby combinations of these values for which the Routh-Hurwitz (RH) stability conditions are satisfied \cite[]{RoHu} for the numerical stability of the model.  We also need to make sure that the $\kappa/g$ ratios are all above the quoted physically achievable minima.  We call this part ``pre-optimization.''  Then we can calibrate the program and be ready to look for regions of entanglement.
\subsubsection{\label{sec:II.B.1}Initial Parameter Values}
Our starting values are summarized in \Table{1}, with unenhanced laser-coupling strength $g_0  = 5.65 \times 10^6 \,[{\textstyle{{\text{rad}} \over \text{s}}}]$.
\begin{table}[H]
\caption{\label{tab:1}Initial parameter values, explained in \App{App.B}.}
\begin{ruledtabular}
\begin{tabular}{c|cc}
Type $k$ & $\kappa_k$ $[{\frac{\text{rad}}{\text{s}}}]$& $g_k$ $[{\frac{\text{rad}}{\text{s}}}]$\\[0.5ex]
\hline
\\[-7pt]
$mw$ & $5.65\times 10^5$ & $5.65\times 10^6$ \\
 $o$ & $3.77\times 10^5$ & $7.53\times 10^5$ \\
 $m$ & $1.88\times 10^3$ & $5.65\times 10^6$ \\
 $f$ & $6.27\times 10^8$ & $2.67\times 10^9$ \\
\end{tabular}
\end{ruledtabular}
\end{table}
From \cite[]{LinT}, the achievable minimum $\kappa/g$ ratios are
\begin{equation}
({\textstyle{{\kappa _{mw} } \over {g_{mw} }}})_{\text{min}}{\kern -1pt}={\kern -1pt} 0.1,\;\;\;({\textstyle{{\kappa _o } \over {g_o }}})_{\text{min}}{\kern -1pt}={\kern -1pt} 0.5,\;\;\;({\textstyle{{\kappa _m } \over {g_m }}})_{\text{min}}{\kern -1pt}={\kern -1pt} 3.33 \times {\kern -2pt}10^{ - 4}{\kern -2pt}.
\label{eq:13}
\end{equation}
These ratios are the minimum amount of damping per driving strength reported as achievable, and represent how well these objects can dissipate energy that is pumped into them without catastrophic failure.
\subsubsection{\label{sec:II.B.2}Routh-Hurwitz Stability Conditions}
From \cite[]{RoHu}, if we recast the Langevin equation as
\begin{equation}
{\textstyle{{d\mathbf{v}(t)} \over {dt}}} = H(t)\mathbf{v}(t) + \sqrt{K} \mathbf{v}_{\text{in}} (t),
\label{eq:14}
\end{equation}
where $H\!\equiv\!-iM$, then the system is Routh-Hurwitz (RH) stable if all the real parts of the eigenvalues of $H$ are negative.  Thus, a good metric for RH stability is
\begin{equation}
S_{\text{RH}}\equiv\text{max}(\text{Re}(\text{eig}(H))),
\label{eq:15}
\end{equation}
where \smash{$S_{\text{RH}}<0$} is the necessary and sufficient condition for RH stability. Note that \smash{$S_{\text{RH}}=0$} is an indeterminate case, which is not necessarily stable or unstable.

Minimizing \smash{$S_{\text{RH}}$} over all parameter combinations is the main objective of our pre-optimization; basically we want to find a possibly new set of initial parameter values that are RH stable.  However, these values \textit{also} need to have physically achievable $\kappa/g$ ratios.
\subsubsection{\label{sec:II.B.3}Physical Parameter Assignment}
Before we can start the pre-optimization, we have to connect the theoretical parameters to the physical parameters.  Since we want to generate two-mode squeezing between the microwave cavities, a first guess for achieving this is to choose the assignment
\begin{equation}
\begin{array}{*{20}l}
   {g_{mw,1}  = g_0 \cosh (r),} & {g_{o,1}  \in [g_0 ,100g_0 ],}  \\
   {g_{mw,2}  = g_0 \sinh (r),} & {g_{o,2}  \in [g_0 ,100g_0 ].}  \\
\end{array}
\label{eq:16}
\end{equation}
Then, we could express the Bogoliubov modes as linear combinations of the cavity modes where the coefficients are proportional to the driving laser strengths as
\begin{equation}
\begin{array}{*{20}l}
   {\beta _1 } &\!\! { = {\textstyle{1 \over {g_0 }}}(g_{mw,1} a_1  + ig_{mw,2} a_2 ^{\dag}  )}  \\
   {\beta _2 ^{\dag}  } &\!\! { = {\textstyle{1 \over {g_0 }}}( - ig_{mw,2} a_1  + g_{mw,1} a_2 ^{\dag} ),}  \\
\end{array}
\label{eq:17}
\end{equation}
where the squeezing parameter is then
\begin{equation}
r = \tanh ^{ - 1} ({\textstyle{{g_{mw,2} } \over {g_{mw,1} }}}).
\label{eq:18}
\end{equation}
Of course many other assignments are possible, so the choice in \Eq{16} may not be the most effective, but it is a convenient starting place for further investigations.
\subsubsection{\label{sec:II.B.4}Choice of Target Squeezing Parameter}
Since the logarithmic negativity $E_\mathcal{N}$ \cite[]{Wern}, the entanglement measure we use in \Sec{II.C}, has range $E_\mathcal{N} \in [0,\infty )$ for infinite-level systems, we need to know what value is ``high enough'' for \textit{near-maximal} entanglement.

From \cite[]{HedD,HedE}, a properly normalized multipartite entanglement monotone is the \textit{ent}, which for a two-mode squeezed state with squeezing parameter $r \in [0,\infty )$ is
\begin{equation}
\Upsilon (r) = 1 - {\textstyle{1 \over {2\cosh ^2 (r) - 1}}} \in [0,1],
\label{eq:19}
\end{equation}
where $\Upsilon(0)=0$ corresponds to the separable two-mode vacuum state, and the value of $\Upsilon(\infty)=1$ is for the maximally entangled two-mode squeezed state; see \App{App.C}.

Though our two cavities are not an isolated system, focusing on them effectively traces over the rest of the system, and \Eq{19} acts as a \textit{mathematical lens} for viewing how good the squeezing parameter would be if the system were isolated and started from vacuum.  Thus, we can specify any \textit{target ent} $\Upsilon_{\ast} \in [0,1]$ that we want, and obtain a \textit{target ideal squeezing parameter} $r_{\ast}$ from
\begin{equation}
r_{\ast} (\Upsilon _{{\ast}})  = \cosh ^{ - 1} \left( {\sqrt {{\textstyle{1 \over 2}}({\textstyle{1 \over {1 - \Upsilon _{\ast} }}} + 1)} } \right).
\label{eq:20}
\end{equation}
For example, if we want an ent of $\Upsilon_{\ast} =0.999$, \Eq{20} yields
\begin{equation}
r_{\ast} (0.999)\approx 3.8,
\label{eq:21}
\end{equation}
which shows that a relatively low value of $r$ is sufficient for near-maximal entanglement of the reduced system of the cavities, assuming that the modes evaluated for entanglement are Bogoliubov modes, even if momentarily.
\subsubsection{\label{sec:II.B.5}Pre-Optimization Results}
Pre-optimization seeks a ``winning'' parameter-value combination $\mathbf{g} \equiv (g_{mw,1} ,g_{mw,2} ,g_{o,1} ,g_{o,2} )$ (where the functional definitions of its elements would change if we choose a different physical parameter assignment than that in \Eq{16}), such that the RH metric of \Eq{15} satisfies $S_{\text{RH}}  < 0$, while the $\kappa/g$ ratios are all kept above the achievable minima in \Eq{13}, and we try to keep squeezing parameter $r$ as high as possible, meaning $r\ge 3.8$, the ``good enough'' value from \Eq{21}.  The purpose of pre-optimization is to show that parameter combinations exist that can satisfy all the desired constraints, which in our case are indepedent of the solution to the Langevin equation (though they can still depend on time).

\begin{figure}[H]
\centering
\includegraphics[width=1.00\linewidth]{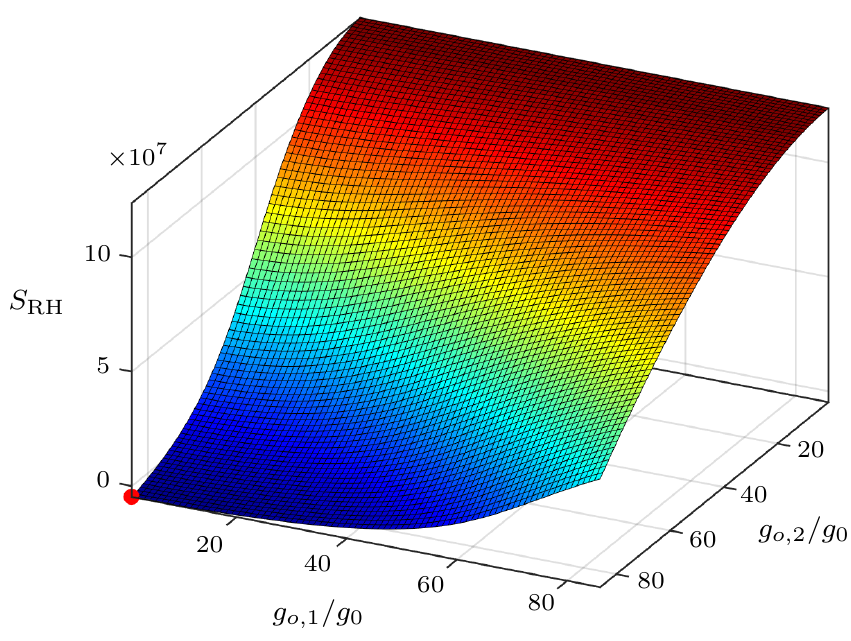}%
\vspace{-6pt}
\caption[]{(color online) Plot of RH metric $S_{\text{RH}}(H)$ from \Eq{15}, where $H\equiv-iM$, where $M$ is given in \Eq{10}, and we use only fixed parameter values here, as mentioned in the text.  The surface shows RH-stability metric $S_{\text{RH}}$ for all parameter combinations tested, and the red dot is the winner $\min(S_{\text{RH}})$.  All nonlaser parameters were held constant, with $r=3.85$.}
\label{fig:2}
\end{figure}

As a first iteration, the values of \Table{1} were used for a static-driving-laser snapshot of the RH stability, using $r=3.8$.  Then, the damping strengths $\kappa$ were increased manually (permissible since that simply implies that we make more heavily damped cavities) while increasing $r$ and checking the $\kappa/g$ ratios against \Eq{13}.  This resulted in new damping values,
\begin{equation}
\kappa _{mw}  = 20C,\;\;\;\kappa _o  = 60C,\;\;\;\kappa _m  = 0.001C,
\label{eq:22}
\end{equation}
where $C$ is given in \Eq{B.5}, and increased squeezing parameter $r=3.85$. The optical damping was then further adjusted as
\begin{equation}
\kappa _{o,1}  = \kappa _o ,\;\;\;\kappa _{o,2}  = 2.5\kappa _o ,
\label{eq:23}
\end{equation}
to keep all $\kappa/g$ ratios above their minima.  \Figure{2} shows the RH metric tested over this parameter set.
The $\kappa/g$ ratios for this parameter set were
\begin{equation}
\begin{array}{*{20}l}
   {{\textstyle{{\kappa _{mw,1} } \over {g_{mw,1} }}} = 0.284,} & {{\textstyle{{\kappa _{o,1} } \over {g_{o,1} }}} = 20.0,} & {{\textstyle{{\kappa _{m,1} } \over {g_{m,1} }}} = 3.33\times 10^{-4},}  \\[8pt]
   {{\textstyle{{\kappa _{mw,2} } \over {g_{mw,2} }}} = 0.284,} & {{\textstyle{{\kappa _{o,2} } \over {g_{o,2} }}} = 0.581,} & {{\textstyle{{\kappa _{f} } \over {g_{f} }}} {\kern 8pt}= 0.236,}  \\
\end{array}
\label{eq:24}
\end{equation}
which shows that we have found \textit{physically achievable} $\kappa/g$ ratios while also attaining the ``good'' squeezing parameter $r=3.85$, and the safely negative RH-metric value,
\begin{equation}
\min (S_{\text{RH}} ) =  - 4.75 \times 10^6  \ll 0.
\label{eq:25}
\end{equation}

Note that finding a valid parameter combination with a ``good'' $r$ does \textit{not} mean that cavities have the entanglement corresponding to $r$; rather it means that \textit{in the reference frame of the Bogoliubov modes}, the system behaves as a two-mode squeezed state with that $r$.  If at any times the time-dependent cavity modes are equal to those Bogoliubov modes, then their entanglement will reach that value at those instances.

In fact, we found that it was possible to construct the Bogoliubov modes from linear combinations of the eigenmodes of $H$, numerically verified to 15 decimal places, even with damping. Therefore, near-maximal entanglement \textit{is possible} in this system.  However, the expressions are highly nonlinear and transcendental, making it infeasible to analytically obtain the conditions for instances of Bogoliubov modes, as was done in \cite[]{LinT}.  Therefore we must use numerical methods here.

Now we are ready to construct an optimization that includes all of these features, including entanglement, which we develop next.
\subsection{\label{sec:II.C}Entanglement Calculation}
As explained in detail in \App{App.D}, the entanglement of the two micrwoave cavities as logarithmic negativity is
\begin{equation}
E_{\mathcal{N}}  = \max \{ 0, - \log _2 (2r_0 )\},
\label{eq:26}
\end{equation}
where
\begin{equation}
r_0  = \sqrt {{\textstyle{{b - \sqrt {b^2  - 4c} } \over 2}}},
\label{eq:27}
\end{equation}
with scalars
\begin{equation}
b = \det (A) + \det (B) - 2\det (C)\;\;\;\text{and}\;\;\;c=\det(V), 
\label{eq:28}
\end{equation}
where $V$ is the covariance matrix,
\begin{equation}
V = \left( {\begin{array}{*{20}c}
   A & C  \\
   {C^T } & B  \\
\end{array}} \right),
\label{eq:29}
\end{equation}
with marginal block-matrices
\begin{equation}
A = (\langle a_1 ^{\dag}  a_1 \rangle  + {\textstyle{1 \over 2}})\left( {\begin{array}{*{20}c}
   1 & 0  \\
   0 & 1  \\
\end{array}} \right)\!,\;\;B = (\langle a_2 ^{\dag}  a_2 \rangle  + {\textstyle{1 \over 2}})\left( {\begin{array}{*{20}c}
   1 & 0  \\
   0 & 1  \\
\end{array}} \right)\!,
\label{eq:30}
\end{equation}
and correlation block matrix,
\begin{equation}
C = \left( {\begin{array}{*{20}r}
   {{\textstyle{1 \over 2}}(\langle a_1 a_2 \rangle  + \langle a_1 ^{\dag}  a_2 ^{\dag}  \rangle )} & {{\textstyle{1 \over {2i}}}(\langle a_1 a_2 \rangle  - \langle a_1 ^{\dag}  a_2 ^{\dag}  \rangle )}  \\
   {{\textstyle{1 \over {2i}}}(\langle a_1 a_2 \rangle  - \langle a_1 ^{\dag}  a_2 ^{\dag}  \rangle )} & { - {\textstyle{1 \over 2}}(\langle a_1 a_2 \rangle  + \langle a_1 ^{\dag}  a_2 ^{\dag}  \rangle )}  \\
\end{array}} \right)\!,
\label{eq:31}
\end{equation}
(not to be confused with the scalar $C$ in \Eq{22}) where the time-dependent expectation-value functions are given in \App{App.D} in (\ref{eq:D.9}--\ref{eq:D.12}).

For reference, the logarithmic negativity in terms of the ideal squeezing parameter $r_{*}$ and ideal ent $\Upsilon_{*}$ is
\begin{equation}
E_{\mathcal{N}}  = {\textstyle{{1 } \over {\ln (\sqrt 2 )}}}r_* = {\textstyle{{1 } \over {\ln (\sqrt 2 )}}}\cosh ^{ - 1}\! \left( {\sqrt {{\textstyle{1 \over 2}}({\textstyle{1 \over {1 - \Upsilon _{\ast} }}} + 1)} } \right),
\label{eq:32}
\end{equation}
where the term \textit{ideal} refers to what these quantities \textit{would mean} if the two microwave cavities were in an isolated two-mode squeezed state \cite{HedE}.
\section{\label{sec:III}Results and Optimization of Entanglement}
Here we present the results of several types of searches for near-maximal entanglement of the driven cavities.  Due to the combinatorially-hard nature of this problem, these results are by no means exhaustive, yet they demonstrate that it should be possible to achieve entanglement as a resource for quantum teleportation between remote microwave cavities with optical fiber connections.
\subsection{\label{sec:III.A}Numerical Optimization of Entanglement}
The highest entanglement found from search-based optimization was achieved by abandoning the notion of hyperbolically fixing two of the driving lasers (as in \Eq{16}), and instead allowing all four driving lasers to be pulse shapes with all parameters allowed to roam within certain domains, tested over many iterations and fueled by a pseudo-random number generator. 

\begin{figure}[H]
\centering
\includegraphics[width=1.00\linewidth]{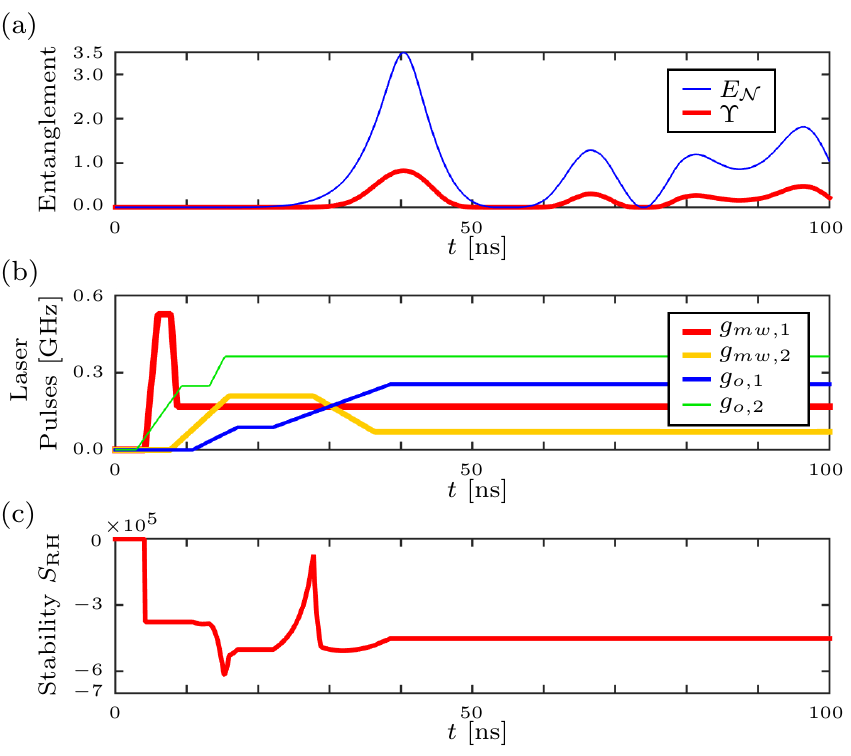}%
\vspace{-6pt}
\caption[]{(color online) Random-search-based winner over sets of four trapezoidal pulses. (a) Entanglement as logarithmic negativity $E_{\mathcal{N}}\in[0,\infty)$ and ent $\Upsilon\in[0,1]$. (b) Driving-laser pulses. (c) RH-stability metric $S_{\text{RH}}$. Trapezoidal-rule step number was $n=5$, and propagator step number was $N=50$ for computational efficiency ($n=10$ and $N=1600$ are best for convergence, but the difference is negligible); see \App{App.F}.  All subsystems were in the vacuum state initially, and $1000$ time points were used, connected by line segments here and in later plots, though keep in mind that may not be the ``truth.''}
\label{fig:3}
\end{figure}

The pulses were doubly-asymmetric trapezoidal functions, where starting height, peak height, and end heights could all be different, as well as start time, rise time, peak width, and fall time, and the starting height was constrained to be zero.  \Figure{3} shows the results of the search for a parameter combination that caused an instant of highest logarithmic negativity such that all time points were RH-stable, and all $\kappa/g$ ratios were within an order of magnitude of the achievable minima.

The peak entanglement in \Fig{3} is
\begin{equation}
\text{max}(E_{\mathcal{N}})=3.49,\;\;\;\text{max}(\Upsilon)=0.823,
\label{eq:33}
\end{equation}
which is comparable to $\text{max}(E_{\mathcal{N}})\approx 3.45$ derived analytically in \cite[]{LinT} for the simpler system of two mechanically coupled cavities.  Here, in the zero-damping limit, we get $\text{max}(\Upsilon)=0.939$, though in that case the stability is indeterminate without the damping to act as ballast.

The damping used in \Fig{3} is
\begin{equation}
\kappa_{mw}=0.8C,\;\;\;\kappa_{o}=0.9C,\;\;\;\kappa_{m}=0.001C,
\label{eq:34}
\end{equation}
however, the search method used the zero-damping case to switch-off unnecessary expressions to pre-screen for quasi-stable results at higher speed, using the low positive threshold $S_{\text{RH}}<10^{-4}$ instead of the proper condition $S_{\text{RH}}<0$. Then the candidate pulse sets were applied to the damped system with the correct stability condition, and the pulse set with the highest peak $E_{\mathcal{N}}$ was the winning set, as seen in \Fig[b]{3}.

\begin{figure}[H]
\centering
\includegraphics[width=1.00\linewidth]{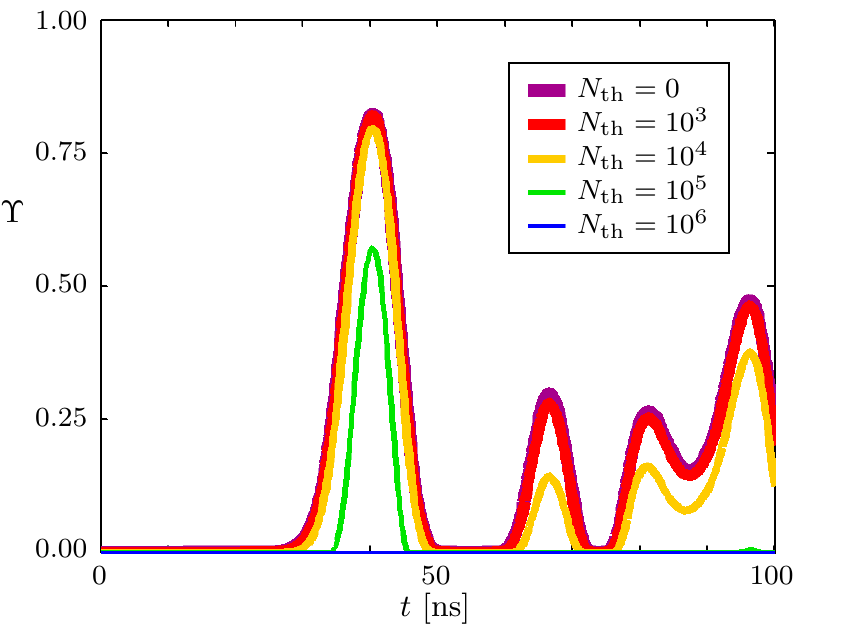}%
\vspace{-2pt}
\caption[]{(color online) Ent values of the damped system from \Fig{3}, shown here for various values of thermal occupation number $N_{\text{th}}$ for the mechanical oscillator baths. The peak ent is $0.823$ for $N_{\text{th}}=0$, and remains near this until $N_{\text{th}}=10^3$, after which it dies to zero near $N_{\text{th}}=10^6$.}
\label{fig:4}
\end{figure}

The minimum $\kappa/g$ ratios achieved in \Fig{3} are
\begin{equation}
\begin{array}{*{20}l}
   {\min ({\textstyle{{\kappa _{mw,1} } \over {g_{mw,1} }}})} &\!\! { = 2.86 \times 10^{ - 3} } &\;\; {\min ({\textstyle{{\kappa _{o,1} } \over {g_{o,1} }}})} &\!\! { = 6.66 \times 10^{ - 3} }  \\[8pt]
   {\min ({\textstyle{{\kappa _{mw,2} } \over {g_{mw,2} }}})} &\!\! { = 7.22 \times 10^{ - 3} } &\;\; {\min ({\textstyle{{\kappa _{o,2} } \over {g_{o,2} }}})} &\!\! { = 4.67 \times 10^{ - 3}, }  \\
\end{array}
\label{eq:35}
\end{equation}
which are two orders of magnitude less than the minimum achievable values of \Eq{13}. However, these mimima only need to be attained briefly ($50\,[\text{ns}]$), and therefore this may be an achievable regime now or in the near future.
So far, we have modeled everything as starting in the vacuum state, but if we model the mechanical oscillators' baths as being initially in states with mean thermal occupation numbers $N_{\text{th}}$, the entanglement drops further, as seen in \Fig{4}.

As \Fig{4} shows, the entanglement of the damped microwave cavities is fairly robust against mechanical oscillator noise up to about $N_{\text{th}}=10^3$.  Note that since stability $S_{\text{RH}}$ depends on $H$, it is independent of $N_{\text{th}}$.
\subsection{\label{sec:III.B}Frequency-Filtered Entanglement Optimization}
An alternative way to study entanglement in this system is to look at the effect of operating the system only in a narrow band of frequencies, meaning that the solution to the Langevin equation is filtered to only allow Fourier components from a narrow frequency band.

Using only constant laser driving strengths, the solution to the frequency-filtered Langevin equation is
\begin{equation}
\mathbf{v}'_{\text{out}} (\omega _n ) = S(\omega _n )\mathbf{v}'_{\text{in}} (\omega _n ),
\label{eq:36}
\end{equation}
where $\omega _n  \equiv n\Delta \omega $ is a discrete set of (angular) frequencies such that $n$ is an integer and $\Delta \omega$ is a narrow frequency window, and $S\equiv S(\omega)$ is the transfer matrix
\begin{equation}
S(\omega) \equiv I - i\sqrt{K} (\omega I - M)^{ - 1} \sqrt{K}.
\label{eq:37}
\end{equation}
The elements of $\mathbf{v}'_{\text{out}}(\omega _n )$ and $\mathbf{v}'_{\text{in}}(\omega _n )$ in \Eq{36} are annihilation and creation operators that are summations over mode operators of all possible frequencies weighted by a band-pass filtering function, as described in \App{App.E}.

\begin{figure}[H]
\centering
\includegraphics[width=1.00\linewidth]{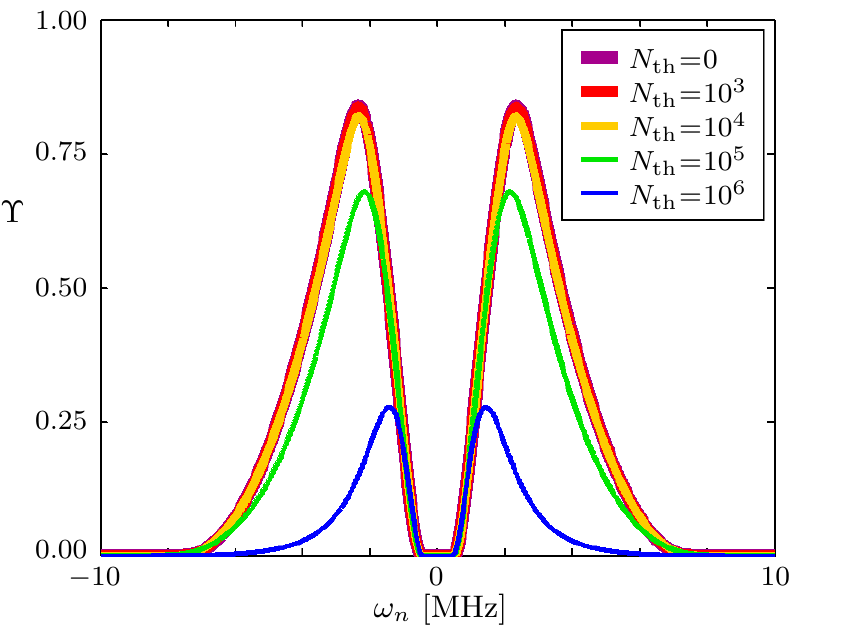}%
\vspace{-2pt}
\caption[]{(color online) Ent values for the frequency-filtered damped system for a range of different angular frequency windows $\omega_n$.  The peak ent is 0.839 for $N_{\text{th}}=0$, and like \Fig{4}, remains near this through $N_{\text{th}}=10^{3}$, but decreases more slowly with $N_{\text{th}}$.  For all curves, $S_{\text{RH}}=-3.80\times 10^{5}$.}
\label{fig:5}
\end{figure}

To attempt to optimize the frequency-filtered solution, we performed a random search of $10^5$ trials producing sets of random driving-laser strengths $g_{k}\in[0,110g_{0}]$, and designated the winning set as the one that produced the largest peak $E_{\mathcal{N}}$ while producing $S_{\text{RH}}<0$ for RH-stability.  The result is shown in \Fig{5}.
The damping in \Fig{5} is that of \Eq{34}, and the winning $g_k$ values that produced \Fig{5} are
\begin{equation}
\begin{array}{*{20}l}
   {g_{mw,1}  = 121.9\;[\text{MHz}],} & {g_{o,1} = 605.4\;[\text{MHz}],}  \\
   {g_{mw,2}  = 105.5\;[\text{MHz}],} & {g_{o,2} = 549.6\;[\text{MHz}],}  \\
\end{array}
\label{eq:38}
\end{equation}
so the $\kappa/g$ ratios are 
\begin{equation}
\begin{array}{*{20}l}
   {{\textstyle{{\kappa _{mw,1} } \over {g_{mw,1} }}}} &\!\! { = 1.234\times 10^{-2}\!, } &\;\; {{\textstyle{{\kappa _{o,1} } \over {g_{o,1} }}}} &\!\! { = 2.795\times 10^{-3}\!, }  \\[8pt]
   {{\textstyle{{\kappa _{mw,2} } \over {g_{mw,2} }}}} &\!\! { = 1.426\times 10^{-2}\!, } &\;\; {{\textstyle{{\kappa _{o,2} } \over {g_{o,2} }}}} &\!\! { = 3.079\times 10^{-3}\!, }  \\
\end{array}
\label{eq:39}
\end{equation}
which are all about one order of magnitude less than the minimum achievable values of \Eq{13}, which again, may be within the realm of possibility.

Thus, frequency-filtering this system can achieve near-maximal entanglement while maintaining RH-stability.
\section{\label{sec:IV}Conclusions}
We have derived and explored the entanglement properties of a nonadiabatic linearized quantum-mechanical model of two laser-driven optomechanical systems (each consisting of one microwave cavity and one optical cavity coupled by a two-sided mirror acting as a mechanical oscillator) coupled by a short single-mode optical fiber connected to the optical cavities.

In particular, we explored this system's ability to achieve and maintain high levels of entanglement between the microwave cavities.  The purpose of this is to enable a continuous-variable quantum teleportation scheme for ideal state transfer between microwave cavities, the intended application being a distributed superconducting-qubit architecture to improve the scalability of superconducting-qubit quantum computers.

Since this system is large enough to contain a computationally difficult number of parameters, we approached the entanglement optimization numerically from several different perspectives.  

First, \Sec{III.A} did a random search of parameters describing a set of four general trapezoidal laser pulses, where the starting height was constrained to zero, but the ending height was allowed to be larger than the central height.  This method also pre-screened its candidates based on their approximate Routh-Hurwitz- (RH)-stability properties as calculated in the zero-damping regime, before evaluating the passing candidates in the damping regime.  This method provided us with the result in \Fig{3} that shows a decent level of entanglement, nearly exactly the same as was found in \cite[]{LinT} analytically, while also maintaining RH-stability.  The downsides to this were that the entanglement only lasts on the order of ns, and it required $\kappa/g$ ratios that were two orders of magnitude lower than known-achievable values.  However, one positive aspect of this example is its robustness against thermal noise, as seen in \Fig{4} showing that the entanglement stays approximately unchanged for mean thermal occupation numbers up to $N_{\text{th}}=10^3$ for the baths of the mechanical oscillators.

Finally, \Sec{III.B} examined a frequency-filtered model, similar to that in \cite[]{LinT}. This method uses constant laser pulses and looks at only a narrow band of operating frequencies of the solution of equation of motion (the Langevin equation).  The method was a random numerical search over the four laser pulse strengths, looking for the set that caused the highest peak entanglement while maintaining RH-stability.  The winning result over $10^5$ trials produced a value only sightly higher than that found in \Sec{III.A}, and the system again showed the same robustness against thermal noise up to $N_{\text{th}}=10^3$.  While the results in \cite[]{LinT} show a much higher peak entanglement, examination of its methods in \cite[]{LSup} shows a disagreement between our derivations, since there, the Langevin equation is first adiabatically approximated but then the input-output relation is omitted, whereas here we did not make the adiabatic approximation and used the input-output relation.  Therefore the qualitative differences in the results may be due to a derivation error in one of these methods, or possibly they arise from using different approximations.

Since the writing of this paper began, several advances have been made in similar areas in \cite{CSYW,APTV,AbTV,HoWu,CeHa}, though each for a slightly different system, and with different approaches.  Recently, others have explored alternative control methods, such as \cite{Stef} where cavity coupling strengths are modulated.  Perhaps similar variations may help future investigations to surpass our results.

Ultimately, the main results of this work are that this kind of interconnected driven quantum system is capable of achieving significant levels of entanglement for fairly decent amounts of time, which may enable teleportation schemes \cite[]{BrLo} to accomplish high-fidelity state transfer between remote microwave cavities connected via optical fibers.  Our results suggest that distributed superconducting quantum computing may be feasible, particularly if damping-to-driving-strength ratios can be lowered.
\begin{acknowledgments}
This work was funded by The Johns Hopkins University Applied Physics Laboratory's Internal Research and Development Program as well as its Postdoctoral Fellowship in Quantum Information Science.
\end{acknowledgments}
\begin{appendix}
\section{\label{sec:App.A}Derivation of the Langevin Equation}
The Heisenberg equations for the bath operators are
\begin{equation}
{\textstyle{{dc_k (\omega ,t)} \over {dt}}} = i[{\textstyle{1 \over \hbar }}H(t),c_k (\omega ,t)] = i[{\textstyle{1 \over \hbar }}H_{dk} (t),c_k (\omega ,t)],
\label{eq:A.1}
\end{equation}
where the bath operators obey
\begin{equation}
[c_j (\omega ,t),c_k ^{\dag}  (\omega ',t)] = \delta _{j,k} \delta (\omega  - \omega ').
\label{eq:A.2}
\end{equation}
Then putting \Eq{6} and \Eq{A.2} into \Eq{A.1} yields
\begin{equation}
i{\textstyle{{dc_k (\omega ,t)} \over {dt}}} = (\omega  - \omega _k )c_k (\omega ,t) - ig_{dk} a_k (t),
\label{eq:A.3}
\end{equation}
where we used the bosonic commutation relations $[a_j ,a_k ^{\dag}  ] = \delta _{j,k}$ for the primary mode operators.  Direct integration of \Eq{A.3} gives solutions
\begin{equation}
\begin{array}{*{20}l}
   {c_k (\omega ,t) = } &\!\! {e^{ - i(\omega  - \omega _k )(t - t_0 )} c_k (\omega ,t_0 )}  \\
   {} &\!\! { - g_{dk} \int_{t_0 }^t {a_k (t')e^{ - i(\omega  - \omega _k )\left( {t - t'} \right)} dt'.} }  \\
\end{array}
\label{eq:A.4}
\end{equation}

The Heisenberg equations for the primary modes are
\begin{equation}
{\textstyle{{da_k (t)} \over {dt}}} = i[{\textstyle{1 \over \hbar }}H(t),a_k (t)],
\label{eq:A.5}
\end{equation}
and putting \Eq{1} into \Eq{A.5} gives
\begin{equation}
{\kern -1pt}\begin{array}{*{20}l}
   {i{\textstyle{{da_{o,1} } \over {dt}}}} &\!\! { = g_{o,1} b_{m,1}  + g_f f + i\int_{ - \infty }^\infty  {} g_{do,1} c_{o,1} (\omega ,t)d\omega }  \\
   {i{\textstyle{{db_{m,1} } \over {dt}}}} &\!\! { = g_{o,1} a_{o,1} \!+\! g_{mw,1} a_{mw,1} \!\!+\! i\!\int_{ - \infty }^\infty  {}\! g_{dm,1} c_{m,1} (\omega ,t)d\omega }  \\
   {i{\textstyle{{da_{mw,1} } \over {dt}}}} &\!\! { = g_{mw,1} b_{m,1}  + i\int_{ - \infty }^\infty  {} g_{dmw,1} c_{mw,1} (\omega ,t)d\omega }  \\
   {i{\textstyle{{df} \over {dt}}}} &\!\! { = g_f a_{o,1}  - g_f a_{o,2}  + i\int_{ - \infty }^\infty  {} g_{df} c_f (\omega ,t)d\omega }  \\
   {i{\textstyle{{da_{o,2} } \over {dt}}}} &\!\! { = g_{o,2} b_{m,2}  - g_f f + i\int_{ - \infty }^\infty  {} g_{do,2} c_{o,2} (\omega ,t)d\omega }  \\
   {i{\textstyle{{db_{m,2} } \over {dt}}}} &\!\! { = g_{o,2} a_{o,2} \!+\! g_{mw,2} a_{mw,2} ^{\dag} \!\!+\! i\!\int_{ - \infty }^\infty  {}\! g_{dm,2} c_{m,2} (\omega ,t)d\omega }  \\
   {i{\textstyle{{da_{mw,2} } \over {dt}}}} &\!\! { = g_{mw,2} b_{m,2} ^{\dag}   + i\int_{ - \infty }^\infty  {} g_{dmw,2} c_{mw,2} (\omega ,t)d\omega .}  \\
\end{array}
\label{eq:A.6}
\end{equation}
Now eliminate the time-dependent bath operators by putting \Eq{A.4} into the integrals in \Eq{A.6} to get 
\begin{equation}
i\int_{ - \infty }^\infty  {} g_{dk} c_k (\omega ,t)d\omega  =  - i{\textstyle{{\kappa _k } \over 2}}a_k  + i\sqrt {{\textstyle{{\kappa _k } \over {2\pi }}}} a_{k,\text{in}},
\label{eq:A.7}
\end{equation}
where we defined ``damping strengths,'' 
\begin{equation}
\kappa _k  \equiv 2\pi g_{dk} ^2\;\;\; \text{so that}\;\;\;g_{dk}  = \sqrt {{\textstyle{{\kappa _k } \over {2\pi }}}},
\label{eq:A.8}
\end{equation}
and ``input operators,''
\begin{equation}
a_{k,\text{in}} (t) \equiv \int_{ - \infty }^\infty  {}\!\! e^{ - i(\omega  - \omega _k )(t - t_0 )} c_k (\omega ,t_0 )d\omega ,
\label{eq:A.9}
\end{equation}
and where we used the facts that
\begin{equation}
\begin{array}{*{20}l}
   {\int_{ - \infty }^\infty  {} e^{ - i(\omega  - \omega _{o,i} )\left( {t - t'} \right)} d\omega } &\!\! { = e^{i\omega _{o,i} \left( {t - t'} \right)} 2\pi \delta (t - t')}  \\
   {\int_{t_0 }^t {f(t')\delta (t - t')dt'} } &\!\! { = {\textstyle{1 \over 2}}f(t),}  \\
\end{array}
\label{eq:A.10}
\end{equation}
where the top result in \Eq{A.10} comes from the nonunitary Fourier transform $\mathcal{F}[1] = \int_{ - \infty }^\infty  {} 1e^{ - i2\pi xf} df = \delta (x)$, and the bottom result comes from using a boxcar model to integrate over a Dirac delta function with one of its arguments shared by a bound of the integral \cite[]{GaCo}.  

Then, noticing that both creation and annihilation operators appear in the system in \Eq{A.6}, we want to take adjoints of certain equations such that the remaining system has each operator appearing with only one particular type of ``daggerness.''  To achieve this in general, start with the first equation that has a lone dagger on the right side, and identifying that as the standard daggerness for that operator, take adjoints of any remaining equations in which it appears and rearrange the negatives.  Thus, putting \Eq{A.7} into \Eq{A.6} and fixing the daggerness yields
\begin{equation}
\begin{array}{*{20}l}
   {i{\textstyle{{da_{o,1} } \over {dt}}}} &\!\! { =  - i\widetilde{a}_{o,1}  + g_{o,1} b_{m,1}  + g_f f + i\widetilde{a}_{o,1,\text{in}} }  \\
   {i{\textstyle{{db_{m,1} } \over {dt}}}} &\!\! { = g_{o,1} a_{o,1}  - i\widetilde{b}_{m,1}  + g_{mw,1} a_{mw,1}  + i\widetilde{b}_{m,1,\text{in}} }  \\
   {i{\textstyle{{da_{mw,1} } \over {dt}}}} &\!\! { = g_{mw,1} b_{m,1}  - i\widetilde{a}_{mw,1}  + i\widetilde{a}_{mw,1,\text{in}} }  \\
   {i{\textstyle{{df} \over {dt}}}} &\!\! { = g_f a_{o,1}  - i\widetilde{f} - g_f a_{o,2}  + i\widetilde{f}_{\text{in}} }  \\
   {i{\textstyle{{da_{o,2} } \over {dt}}}} &\!\! { =  - g_f f - i\widetilde{a}_{o,2}  + g_{o,2} b_{m,2}  + i\widetilde{a}_{o,2,\text{in}} }  \\
   {i{\textstyle{{db_{m,2} } \over {dt}}}} &\!\! { = g_{o,2} a_{o,2}  - i\widetilde{b}_{m,2}  + g_{mw,2} a_{mw,2} ^{\dag}   + i\widetilde{b}_{m,2,\text{in}} }  \\
   {i{\textstyle{{da_{mw,2} ^{\dag}  } \over {dt}}}} &\!\! { =  - g_{mw,2} b_{m,2}  - i\widetilde{a}_{mw,2} ^{\dag}   + i\widetilde{a}_{mw,2,\text{in}} ^{\dag} , }  \\
\end{array}
\label{eq:A.11}
\end{equation}
where $\widetilde{a}_k \equiv {\textstyle{{\kappa _k } \over 2}}a_k$ and $\widetilde{a}_{k,\text{in}}  \equiv \sqrt {{\textstyle{{\kappa _k } \over {2\pi }}}} a_{k,\text{in}}$.  Thus \Eq{A.11} motivates the definitions of $\mathbf{v}$ and $\mathbf{v}_{\text{in}}$ in \Eq{8}, as well as \Eq{9} and \Eq{10}, and generates the Langevin equation in \Eq{7}.
\section{\label{sec:App.B}Determination of Parameter Values}
From \cite[]{Pell,SeMB}, the single-mode fiber coupling strength is
\begin{equation}
g_f  \approx {\textstyle{{2\pi c\sqrt 2 } \over l}} \approx 2.67 \times 10^9 [{\textstyle{{\text{rad}} \over \text{s}}}],
\label{eq:B.1}
\end{equation}
where we used length $l=1[\text{m}]$, and $c$ is the speed of light in vacuum.  A reasonable fiber damping strength proposed in \cite[]{Pell} was
\begin{equation}
\kappa _f  = 2\pi (100[\text{MHz}]) \approx 6.27 \times 10^8 [{\textstyle{{\text{rad}} \over \text{s}}}],
\label{eq:B.2}
\end{equation}
which yields a $\kappa/g$ ratio of \smash{${\textstyle{{\kappa _f } \over {g_f }}} = 0.236$}.  From \cite[]{AsKM}, a value for unenhanced laser coupling strength was
\begin{equation}
g_0  = 2\pi (9 \times 10^5 [\text{Hz}]) \approx 5.65 \times 10^6 [{\textstyle{{\text{rad}} \over \text{s}}}],
\label{eq:B.3}
\end{equation}
and \cite[]{LinT} quoted the minimum physically achievable ratios given in \Eq{13}, which are not fundamental minima, but rather just what is known to have been achieved.  The minimum $\kappa/g$ ratios represent how much damping is needed to dissipate the energy input of the driving lasers; ratio values lower than those result in catastrophic failure of the cavities, meaning they are permanently ruined. 

Other values given in \cite[]{LinT} are 
\begin{equation}
\kappa _{mw} ^L  = 0.3,\;\;\;\kappa _o ^L  = 0.2,\;\;\;\kappa _m ^L  = 0.001,\;\;\;g_0 ^L  = 3,
\label{eq:B.4}
\end{equation}
all in undefined dimensionless units, indicated by the superscript $L$.  From the information in \Eq{B.4}, we can deduce a conversion factor for our units. Assuming all coupling strengths are proportional to the same $g_0$, we have the proportionality relation $g_k  = Cg_k ^L$, so that
\begin{equation}
C = {\textstyle{{g_0 } \over {g_0 ^L }}} = {\textstyle{{5.65 \times 10^6 [{\textstyle{{\text{rad}} \over \text{s}}}]} \over 3}} \approx 1.88 \times 10^6 [{\textstyle{{\text{rad}} \over \text{s}}}].
\label{eq:B.5}
\end{equation}
Then, using the fact that the ratios are the same for both dimensions, we can write
\begin{equation}
{\textstyle{{\kappa _k } \over {g_k }}} = {\textstyle{{\kappa _k ^L } \over {g_k ^L }}},
\label{eq:B.6}
\end{equation}
which we can solve for the raw damping strengths as
\begin{equation}
\kappa _k  = g_k {\textstyle{{\kappa _k ^L } \over {g_k ^L }}} = g_k {\textstyle{{\kappa _k ^L } \over {g_k /C}}} = \kappa _k ^L C,
\label{eq:B.7}
\end{equation}
which yields the $\kappa_k$ column of \Table{1}.  Putting \Eq{B.7} into \Eq{B.6}, we can obtain physically reachable values for the coupling strengths as
\begin{equation}
g_k  = \kappa _k \left( {{\textstyle{{\kappa _k ^L } \over {g_k ^L }}}} \right)^{ - 1},
\label{eq:B.8}
\end{equation}
which gives the $g_k$ column of \Table{1}, and shows that \cite[]{LinT} used $g_m  = g_0$.  Note that we have supposed that each pair of cavities of a certain type (such as both microwave cavities) share the same damping and coupling strength properties.  Thus, \Eq{B.7} and \Eq{B.8} give us a set of physically reasonable starting values.

To relate the laser coupling strengths to coherent-state parameter $\alpha$, where $|\alpha|^2$ is the mean photon number, (using the excellent assumption that a laser field is in a coherent state) we use the fact from \cite[]{Clad} that
\begin{equation}
g = g_0 \overline{\alpha},
\label{eq:B.9}
\end{equation}
which, using the values from \Table{1}, yields
\begin{equation}
\overline{\alpha}_{mw}  = {\textstyle{{g_{mw} } \over {g_0 }}} = 1,\;\;\;\text{and}\;\;\;\overline{\alpha}_o  = {\textstyle{{g_o } \over {g_0 }}} = 0.133.
\label{eq:B.10}
\end{equation}
Then, since the probability of an ideal on-off detector getting a click from a coherent state input is
\begin{equation}
p_k  = 1 - e^{ - |\overline{\alpha} _k |^2 },
\label{eq:B.11}
\end{equation}
the nonvacuum count probabilities given these values are
\begin{equation}
p_{mw}  = 0.632\;\;\;\text{and}\;\;\;p_o  = 0.0175,
\label{eq:B.12}
\end{equation}
which are both reasonable values to measure given that most thermo-electrically cooled single-photon detectors can yield these values accurately with binomial experiments and good attenuation techniques.
\section{\label{sec:App.C}Ent: A Multipartite Entanglement Monotone}
First presented in \cite[]{HedD}, the \textit{ent} is a multipartite entanglement monotone. For pure-state input, the ent is a measure of how mixed each of the extreme unipartite reductions is for the full multipartite state, meaning that it looks at each of the smallest subsystems over which the multipartite entanglement problem is specified and essentially sums the purities of those subsystems in a way that ensures state-free entanglement normalization.  Like all entanglement monotones, the pure-input ent can be adapted to mixed states via convex-roof extension. See \cite[]{HedE} for details and proofs.
\section{\label{sec:App.D}Entanglement Calculation Details}
\cite[]{LSup} gives the logarithmic negativity as defined in (\ref{eq:26}--\ref{eq:31}), valid for bipartite Gaussian systems, where $V$ is the \textit{quantum covariance matrix of the quadrature variables}, defined in general in \cite[]{Buon} as having elements
\begin{equation}
V_{j,k}  \equiv {\textstyle{1 \over 2}}(\langle u_j u_k \rangle  + \langle u_k u_j \rangle ) - \langle u_j \rangle \langle u_k \rangle ,
\label{eq:D.1}
\end{equation}
where $u_k$ are operators and elements of 
\begin{equation}
\mathbf{u} \equiv (x_1 ,p_1 ,x_2 ,p_2 )^T ,
\label{eq:D.2}
\end{equation}
which is a vector of quadrature operators,
\begin{equation}
x_l  \equiv {\textstyle{1 \over {\sqrt 2 }}}(a_l  + a_l ^{\dag}  )\;\;\;\text{and}\;\;\;p_l  \equiv {\textstyle{1 \over {i\sqrt 2 }}}(a_l  - a_l ^{\dag}  ),
\label{eq:D.3}
\end{equation}
where subscripts $1$ and $2$ in \Eq{D.2} are generic labels of the two modes of the bipartite system.

In our system, since we use the Heisenberg picture, the state of all expectation values is the initial $14$-mode state, and since all subsystems will start in diagonal states, expectation values of mode operators (annihilation or creation operators) will be zero, so the second term in \Eq{D.1} vanishes and we can simply write
\begin{equation}
V = {\textstyle{1 \over 2}}(\langle \mathbf{u}\mathbf{u}^T \rangle  + \langle \mathbf{u}\mathbf{u}^T \rangle ^T ),
\label{eq:D.4}
\end{equation}
as a \textit{special case} of the more general covariance matrix.  Then, putting \Eq{D.3} into \Eq{D.2} and putting that into \Eq{D.4} gives a $V$ that yields (using \Eq{29}),
\begin{equation}
A =\! \left(\!\! {\begin{array}{*{20}c}
   {{\textstyle{1 \over 2}}\!\left(\!\! \begin{array}{l}
 2\langle a_1 ^{\dag}  a_1 \rangle  + 1 \\ 
  + \langle a_1 a_1 \rangle  + \langle a_1 ^{\dag}  a_1 ^{\dag}  \rangle  \\ 
 \end{array}\!\! \right)} &\!\! {{\textstyle{1 \over {2i}}}\!\left( {\langle a_1 a_1 \rangle  - \langle a_1 ^{\dag}  a_1 ^{\dag}  \rangle } \right)}  \\
   {{\textstyle{1 \over {2i}}}\!\left( {\langle a_1 a_1 \rangle  - \langle a_1 ^ {\dag}  a_1 ^{\dag}  \rangle } \right)} &\!\! {{\textstyle{1 \over 2}}\!\left(\!\! \begin{array}{l}
 2\langle a_1 ^{\dag}  a_1 \rangle  + 1 \\ 
  - \langle a_1 a_1 \rangle  - \langle a_1 ^{\dag}  a_1 ^{\dag}  \rangle  \\ 
 \end{array}\!\! \right)}  \\
\end{array}}\!\!\! \right)\!,
\label{eq:D.5}
\end{equation}
\begin{equation}
B =\! \left(\!\! {\begin{array}{*{20}c}
   {{\textstyle{1 \over 2}}\!\left(\!\! \begin{array}{l}
 2\langle a_2 ^{\dag}  a_2 \rangle  + 1 \\ 
  + \langle a_2 a_2 \rangle  + \langle a_2 ^{\dag}  a_2 ^{\dag}  \rangle  \\ 
 \end{array}\!\! \right)} &\!\! {{\textstyle{1 \over {2i}}}\!\left( {\langle a_2 a_2 \rangle  - \langle a_2 ^{\dag}  a_2 ^{\dag}  \rangle } \right)}  \\
   {{\textstyle{1 \over {2i}}}\!\left( {\langle a_2 a_2 \rangle  - \langle a_2 ^ {\dag}  a_2 ^{\dag}  \rangle } \right)} &\!\! {{\textstyle{1 \over 2}}\!\left(\!\! \begin{array}{l}
 2\langle a_2 ^{\dag}  a_2 \rangle  + 1 \\ 
  - \langle a_2 a_2 \rangle  - \langle a_2 ^{\dag}  a_2 ^{\dag}  \rangle  \\ 
 \end{array}\!\! \right)}  \\
\end{array}}\!\!\! \right)\!,
\label{eq:D.6}
\end{equation}
\begin{equation}
C =\! \left(\!\! {\begin{array}{*{20}c}
   {{\textstyle{1 \over 2}}\!\left(\!\! \begin{array}{l}
 \langle a_1 a_2 \rangle  + \langle a_1 a_2 ^{\dag}  \rangle  \\ 
  + \langle a_1 ^{\dag}  a_2 \rangle  + \langle a_1 ^{\dag}  a_2 ^{\dag}  \rangle  \\ 
 \end{array}\!\! \right)} &\!\! {{\textstyle{1 \over {2i}}}\!\left(\!\! \begin{array}{l}
 \langle a_1 a_2 \rangle  - \langle a_1 a_2 ^{\dag}  \rangle  \\ 
  + \langle a_1 ^{\dag}  a_2 \rangle  - \langle a_1 ^{\dag}  a_2 ^{\dag}  \rangle  \\ 
 \end{array}\!\! \right)}  \\
   {{\textstyle{1 \over {2i}}}\!\left(\!\! \begin{array}{l}
 \langle a_1 a_2 \rangle  + \langle a_1 a_2 ^{\dag}  \rangle  \\ 
  - \langle a_1 ^{\dag}  a_2 \rangle  - \langle a_1 ^{\dag}  a_2 ^{\dag}  \rangle  \\ 
 \end{array}\!\! \right)} &\!\! { - {\textstyle{1 \over 2}}\!\left(\!\! \begin{array}{l}
 \langle a_1 a_2 \rangle  - \langle a_1 a_2 ^{\dag}  \rangle  \\ 
  - \langle a_1 ^{\dag}  a_2 \rangle  + \langle a_1 ^{\dag}  a_2 ^{\dag}  \rangle  \\ 
 \end{array}\!\! \right)}  \\
\end{array}}\!\!\! \right)\!,
\label{eq:D.7}
\end{equation}
where we used the Bosonic commutations \smash{$a_l a_l ^{\dag}   = a_l ^{\dag}  a_l  + 1$} to simplify wherever possible.  

Next, we need to find all the expectation values appearing above.  The particular mode solutions of the Langevin equation \Eq{11} for the two modes of interest to us, defining $a_1  \equiv a_{mw,1} (t) = [\mathbf{v}(t)]_3$ and $a_2^{\dag}  \equiv a_{mw,2} ^{\dag}  (t) = [\mathbf{v}(t)]_7$, are
\begin{equation}
\begin{array}{*{20}l}
   {a_1 } &\!\!  =  &\!\! {\sum\limits_{k = 1}^7 \tau _{3,k} (t,0)[\mathbf{v}(0)]_k }  \\
   {} &\!\! {} &\!\! { + \sum\limits_{k = 1}^7 \int_0^t {\tau _{3,k} (t,t')\sqrt {K_{k,k} } [\mathbf{v}_{\text{in}} (t')]_k } dt'}  \\
   {a_2 ^{\dag}  } &\!\!  =  &\!\! {\sum\limits_{k = 1}^7 \tau _{7,k} (t,0)[\mathbf{v}(0)]_k }  \\
   {} &\!\! {} &\!\! { + \sum\limits_{k = 1}^7 \int_0^t {\tau _{7,k} (t,t')\sqrt {K_{k,k} } [\mathbf{v}_{\text{in}} (t')]_k } dt'},  \\
\end{array}
\label{eq:D.8}
\end{equation}
where the labels on the left are the generic labels of the entanglement calculation and the labels on the right are the numerical labels implied by \Eq{8}, where the relevant quantities are defined in \Eq{8}, \Eq{9}, and \Eq{12}.  

Then, forming products of the modes in \Eq{D.8} and their adjoints and taking expectation values leads to
\begin{equation}
\begin{array}{*{20}l}
   {\langle a_1 ^{\dag}  a_1 \rangle  = } &\!\! {|\tau _{3,7} (t,0)|^2 (\overline{n}_{\text{th}}^{(7)}  + 1) + \sum\limits_{k = 1}^6 |\tau _{3,k} (t,0)|^2 \overline{n}_{\text{th}}^{(k)} }  \\
   {} &\!\! { + K_{7,7} (Q_{\text{th}}  + 1)\int_0^t {} |\tau _{3,7} (t,t')|^2 dt'}  \\
   {} &\!\! { + N_{\text{th}} \sum\limits_{k = 2,6} K_{k,k} \int_0^t {} |\tau _{3,k} (t,t')|^2 dt'}  \\
   {} &\!\! { + Q_{\text{th}} \!\!\!\!\sum\limits_{k = 1,3,4,5}\!\!\!\! K_{k,k} \int_0^t {} |\tau _{3,k} (t,t')|^2 dt'},  \\
\end{array}
\label{eq:D.9}
\end{equation}
\begin{equation}
\begin{array}{*{20}l}
   {\langle a_2 ^{\dag}  a_2 \rangle =} &\!\! {|\tau _{7,7} (t,0)|^2 \overline{n}_{\text{th}}^{(7)}  + \sum\limits_{k = 1}^6 |\tau _{7,k} (t,0)|^2 (\overline{n}_{\text{th}}^{(k)}  + 1)}  \\
   {} &\!\! { + K_{7,7} Q_{\text{th}} \int_0^t {|\tau _{7,7} (t,t')} |^2 dt'}  \\
   {} &\!\! { + (N_{\text{th}}  + 1)\sum\limits_{k = 2,6} K_{k,k} \int_0^t {|\tau _{7,k} (t,t')} |^2 dt'}  \\
   {} &\!\! { + (Q_{\text{th}}  + 1)\!\!\!\!\sum\limits_{k = 1,3,4,5}\!\!\!\! K_{k,k} \int_0^t {|\tau _{7,k} (t,t')} |^2 dt'},  \\
\end{array}
\label{eq:D.10}
\end{equation}
\begin{equation}
\begin{array}{*{20}l}
   {\langle a_1 a_2 \rangle  = } &\!\! {\tau _{3,7} (t,0)\tau _{7,7} ^* (t,0)\overline{n}_{\text{th}}^{(7)} }  \\
   {} &\!\! {+ \sum\limits_{k = 1}^6 \tau _{3,k} (t,0)\tau _{7,k} ^* (t,0)(\overline{n}_{\text{th}}^{(k)}  + 1)}\\
   {} &\!\! { + K_{7,7} Q_{\text{th}} \int_0^t {} \tau _{3,7} (t,t')\tau _{7,7} ^* (t,t')dt'}  \\
   {} &\!\! { + (N_{\text{th}}  + 1)\sum\limits_{k = 2,6} K_{k,k} \int_0^t {} \tau _{3,k} (t,t')\tau _{7,k} ^* (t,t')dt'}  \\
   {} &\!\! { + (Q_{\text{th}}  + 1)\!\!\!\!\sum\limits_{k = 1,3,4,5}\!\!\!\! K_{k,k} \int_0^t {} \tau _{3,k} (t,t')\tau _{7,k} ^* (t,t')dt'},  \\
\end{array}
\label{eq:D.11}
\end{equation}
\begin{equation}
\begin{array}{*{20}l}
   {\langle a_1 ^{\dag}  a_2 ^{\dag}  \rangle  = } &\!\! {\tau _{3,7} ^* (t,0)\tau _{7,7} (t,0)(\overline{n}_{\text{th}}^{(7)}  + 1)}  \\
   {} &\!\! { + \sum\limits_{k = 1}^6 \tau _{3,k} ^* (t,0)\tau _{7,k} (t,0)\overline{n}_{\text{th}}^{(k)} }  \\
   {} &\!\! { + K_{7,7} (Q_{\text{th}}  + 1)\int_0^t {} \tau _{3,7} ^* (t,t')\tau _{7,7} (t,t')dt'}  \\
   {} &\!\! { + N_{\text{th}} \sum\limits_{k = 2,6} K_{k,k} \int_0^t {} \tau _{3,k} ^* (t,t')\tau _{7,k} (t,t')dt'}  \\
   {} &\!\! { + Q_{\text{th}} \!\!\!\!\sum\limits_{k = 1,3,4,5}\!\!\!\! K_{k,k} \int_0^t {} \tau _{3,k} ^* (t,t')\tau _{7,k} (t,t')dt'},  \\
\end{array}
\label{eq:D.12}
\end{equation}
where in our model, we let the \textit{primary} subsystems (fiber, cavities, and mechanical oscillators) be initially modeled as thermal states of mean photon number
\begin{equation}
\overline{n}_{\text{th}}^{(m)} \equiv \text{tr}(\rho _{\text{th}}^{(m)} a_{m}^{\dag}  a_{m}) = {\textstyle{1 \over {e^{\hbar \omega _m /k_B T_m }  - 1}}},
\label{eq:D.13}
\end{equation}
for primary subsystem $m$, where $k_B$ is Boltzmann's constant, $T_m$ is the temperature of mode $m$ in kelvins, $\omega _m$ is the field's center frequency, and where we assumed that the baths (the \textit{secondary} subsystems) were initially in diagonal states (in the Fock basis) satisfying
\begin{equation}
{\!\!\!}\begin{array}{*{20}l}
   {\langle a_{j,\text{in}} ^{\dag}  (s')a_{k,\text{in}} (t')\rangle } &\!\! { = Q_{\text{th}} \delta _{j,k} \delta (s' \!-\! t')|_{j,k = 7}}  \\
   {\langle a_{j,\text{in}} (s')a_{k,\text{in}} ^{\dag}  (t')\rangle } &\!\! { = (N_{\text{th}} \! +\! 1)\delta _{j,k} \delta (s' \!-\! t')|_{j,k \in \{ 2,6\}}}  \\
   {\langle a_{j,\text{in}} (s')a_{k,\text{in}} ^{\dag}  (t')\rangle } &\!\! { = (Q_{\text{th}} \! +\! 1)\delta _{j,k} \delta (s' \!-\! t')|_{j,k \in \{ 1,3,4,5\}},}  \\
\end{array}\!\!
\label{eq:D.14}
\end{equation}
\begin{equation}
\begin{array}{*{20}l}
   {\langle a_{j,in} (s')a_{k,\text{in}} ^{\dag}  (t')\rangle } &\!\! { = (Q_{\text{th}} \! +\! 1)\delta _{j,k} \delta (s'\! -\! t')|_{j,k = 7}}  \\
   {\langle a_{j,\text{in}} ^{\dag}  (s')a_{k,\text{in}} (t')\rangle } &\!\! { = N_{\text{th}} \delta _{j,k} \delta (s'\! -\! t')|_{j,k \in \{ 2,6\}}}  \\
   {\langle a_{j,\text{in}} ^{\dag}  (s')a_{k,\text{in}} (t')\rangle } &\!\! { = Q_{\text{th}} \delta _{j,k} \delta (s'\! -\! t')|_{j,k \in \{ 1,3,4,5\}},}  \\
\end{array}
\label{eq:D.15}
\end{equation}
where $N_{\text{th}}$ is the mean thermal occupation number of the initial bath state for the mechanical oscillators, and $Q_{\text{th}}$ is the same quantity but for the baths of all other subsystems, and \smash{$a_{k\in\{2,6\},\text{in}}\equiv b_{m,k\in\{1,2\},\text{in}}$} are the mode operators of the mechanical oscillator baths, and the remaining expectation values are all zero; \smash{$\langle a_1 a_1 \rangle  = 0$}, \smash{$\langle a_1 ^{\dag}  a_1 ^{\dag}  \rangle  = 0$}, \smash{$\langle a_2 a_2 \rangle  = 0$}, \smash{$\langle a_2 ^{\dag}  a_2 ^{\dag}  \rangle  = 0$}, \smash{$\langle a_1 a_2 ^{\dag}  \rangle  = 0$}, \smash{$\langle a_1 ^{\dag}  a_2 \rangle  = 0$}, so that $A$, $B$, and $C$ simplify to \Eq{30} and \Eq{31}.  The integrals needed to compute the expectation values in (\ref{eq:D.9}--\ref{eq:D.12}) were done using the methods given in \App{App.F}.
\section{\label{sec:App.E}Frequency-Filtering Details}
Starting with the Langevin equation from \Eq{7}, expand the time-dependent solutions as
\begin{equation}
\mathbf{v}(t) = {\textstyle{1 \over {\sqrt {2\pi } }}}\int_{ - \infty }^{ \infty } {\mathbf{v}(\omega )e^{ - i\omega t} d\omega },
\label{eq:E.1}
\end{equation}
implying the phasor relation $\frac{d}{dt} \to  - i\omega $, so \Eq{7} becomes
\begin{equation}
\mathbf{v}(t) = i(\omega I - M(t))^{ - 1} \sqrt{K} \mathbf{v}_{\text{in}} (t).
\label{eq:E.2}
\end{equation}
Then, putting \Eq{E.1} into \Eq{E.2} and using the inverse Fourier transform, taking $M$ (and thus driving functions $g_k$) to be constant in time, we get
\begin{equation}
\mathbf{v}(\omega ) = i(\omega I - M)^{ - 1} \sqrt{K} \mathbf{v}_{\text{in}} (\omega ),
\label{eq:E.3}
\end{equation}
where \smash{$\mathbf{v}(\omega )\! =\! {\textstyle{1 \over {\sqrt {2\pi } }}}\int_{ - \infty }^{ \infty } \!{\mathbf{v}(t)e^{i\omega t} dt} $}.  Then, using the input-output relation \smash{$a_{\text{out}} = a_{\text{in}}  - \sqrt{K} a$} \cite[]{AsKM} adapted as
\begin{equation}
\mathbf{v}_{\text{out}} (\omega ) = \mathbf{v}_{\text{in}} (\omega ) - \sqrt{K} \mathbf{v}(\omega ),
\label{eq:E.4}
\end{equation}
putting \Eq{E.3} into \Eq{E.4} gives
\begin{equation}
\mathbf{v}_{\text{out}} (\omega )\! =\! ( {I\! -\! i\sqrt{K} (\omega I\! -\! M)^{ - 1}\! \sqrt{K} })\mathbf{v}_{\text{in}} (\omega ) \!\equiv\! S(\omega )\mathbf{v}_{\text{in}} (\omega ).
\label{eq:E.5}
\end{equation}
Note that the original input-output relation from \cite[]{GaZo} has no minus signs, but sources like \cite[]{AsKM} and \cite[]{LinT} \textit{do} have a minus sign as in \Eq{E.4}.  As mentioned in \Sec{II}, this is because, as in \cite[]{LinT,L1Sp}, the roles of the letters representing mode operators in the dissipation-Hamiltonian are reversed from those in \cite[]{GaCo} upon which \cite[]{GaZo} is based, and the effect is to introduce a sign flip to $g_{dk}$ and hence $\sqrt{K}$ since $g_{dk}=\text{diag}(\sqrt{K})_k$ from \Eq{A.8} and \Eq{9}.  This is merely an unimportant difference in mode-labeling convention.

Then, defining band-filtered mode operators,
\begin{equation}
a'_{k,\text{in/out}} (\omega _n ) \equiv \int_{ - \infty }^{ \infty } {g_D (\omega  - \omega _n )a_{k,\text{in/out}} } (\omega )d\omega ,
\label{eq:E.6}
\end{equation}
where $\omega _n  \equiv n\Delta \omega$ as in \Eq{36}, and square filtering function
\begin{equation}
g_D (\omega ) \equiv \left\{ {\begin{array}{*{20}c}
   {{\textstyle{1 \over {\sqrt {\Delta \omega } }}};} & { - {\textstyle{{\Delta \omega } \over 2}} \le \omega  \le {\textstyle{{\Delta \omega } \over 2}}}  \\
   {0;} & \text{else}  \\
\end{array}} \right.,
\label{eq:E.7}
\end{equation}
we then obtain the frequency-filtered Langevin solutions,
\begin{equation}
\mathbf{v}'_{\text{out}} (\omega _n ) = S(\omega _n )\mathbf{v}'_{\text{in}} (\omega _n ).
\label{eq:E.8}
\end{equation}

The frequency-filtered entanglement of the microwave cavities is then given by putting (\ref{eq:D.5}--\ref{eq:D.7}) into the covariance matrix of \Eq{29} as
\begin{equation}
V(\omega _n ) = \left( {\begin{array}{*{20}c}
   {A(\omega _n )} & {C(\omega _n )}  \\
   {C^T (\omega _n )} & {B(\omega _n )}  \\
\end{array}} \right),
\label{eq:E.9}
\end{equation}
where the operators in (\ref{eq:D.5}--\ref{eq:D.7}) are redefined as \smash{$a'_1 (\omega_n ) \equiv a'_{mw,1,\text{out}} (\omega _n ) = [\mathbf{v}_{\text{out}}^{\prime} (\omega _n )]_3$} and \smash{$a_2^{\prime\dag} (\omega_n )\equiv$} \smash{$ a_{mw,2,\text{out}}^{\prime\dag} (\omega _n ) = [\mathbf{v}_{\text{out}}^{\prime} (\omega _n )]_7$}, which from \Eq{E.8} are
\begin{equation}
\begin{array}{*{20}l}
   {a'_1 (\omega _n )} &\!\! { = \sum\limits_{k = 1}^7 S_{3,k} (\omega _n )[\mathbf{v}_{\text{in}}^{\prime} (\omega _n )]_k }  \\
   {a_2 ^{\prime\dag}  (\omega _n )} &\!\! { = \sum\limits_{k = 1}^7 S_{7,k} (\omega _n )[\mathbf{v}_{\text{in}}^{\prime} (\omega _n )]_k },  \\
\end{array}
\label{eq:E.10}
\end{equation}
where as in \Eq{D.8}, the indices on the left are generic labels for two systems of the entanglement calculation, and the indices on the right correspond to the subsystem labels of the Langevin solution vector.

Again we restrict the initial state of the full system to be a product of diagonal states in the Fock basis so that expectation values of powers of annihilation or creation operators vanish in the Heisenberg picture, as described in \App{App.D}.  The expectation values we need are 
\begin{equation}
\begin{array}{*{20}l}
   {\langle [\mathbf{v}'_{\text{in}} (\omega _n )]_j [\mathbf{v}_{\text{in}} ^{\prime\dag}  (\omega _n )]_j \rangle |_{j \in \{ 1,3,4,5\} } } &\!\! { = Q_{\text{th}}  + 1}  \\
   {\langle [\mathbf{v}_{\text{in}} ^{\prime\dag}  (\omega _n )]_j [\mathbf{v}'_{\text{in}} (\omega _n )]_j \rangle |_{j \in \{ 1,3,4,5\} } } &\!\! { = Q_{\text{th}} }  \\
   {\langle [\mathbf{v}'_{\text{in}} (\omega _n )]_j [\mathbf{v}_{\text{in}} ^{\prime\dag}  (\omega _n )]_j \rangle |_{j \in \{ 2,6\} } } &\!\! { = N_{\text{th}}  + 1}  \\
   {\langle [\mathbf{v}_{\text{in}} ^{\prime\dag}  (\omega _n )]_j [\mathbf{v}'_{\text{in}} (\omega _n )]_j \rangle |_{j \in \{ 2,6\} } } &\!\! { = N_{\text{th}} }  \\
   {\langle [\mathbf{v}'_{\text{in}} (\omega _n )]_7 [\mathbf{v}_{\text{in}} ^{\prime\dag} (\omega _n )]_7 \rangle } &\!\! { = Q_{\text{th}} }  \\
   {\langle [\mathbf{v}_{\text{in}} ^{\prime\dag}  (\omega _n )]_7 [\mathbf{v}'_{\text{in}} (\omega _n )]_7 \rangle } &\!\! { = Q_{\text{th}}  + 1,}  \\
\end{array}
\label{eq:E.11}
\end{equation}
where $N_{\text{th}}$ is the mean thermal occupation number of the initial bath state for the mechanical oscillators, and $Q_{\text{th}}$ is that for the baths of all other subsystems.

Using \Eq{E.11} lets us calculate the expectation values of all mode products, the nonzero results of which are
\begin{equation}
{\!}\begin{array}{*{20}l}
   {\langle a_{1} ^{\prime\dag}  (\omega _n )a'_{1} (\omega _n )\rangle\!= } &\!\! {Q_{\text{th}}\!\!\!\!\!\!\! \sum\limits_{k = 1,3,4,5}\!\!\!\!\!\!\! |S_{3,k} (\omega _n )|^2 \!+\! N_{\text{th}}\!\!\! \sum\limits_{k = 2,6}\!\!\! |S_{3,k} (\omega _n )|^2 }  \\
   {} &\!\! { + (Q_{\text{th}}  + 1)|S_{3,7} (\omega _n )|^2 ,}  \\
\end{array}\!\!
\label{eq:E.12}
\end{equation}
\begin{equation}
\begin{array}{*{20}l}
   {\langle a_2 ^{\prime\dag}  (\omega _n )a'_2 (\omega _n )\rangle\!= } &\!\! { (Q_{\text{th}}  + 1)\!\!\!\!\!\!\sum\limits_{k = 1,3,4,5}\!\!\!\!\!\!\! |S_{7,k} (\omega _n )|^2} \\
   {} &\!\! {+ (N_{\text{th}}  + 1)\!\!\!\sum\limits_{k = 2,6}\!\! |S_{7,k} (\omega _n )|^2 }\\
{} &\!\! {+Q_{\text{th}} |S_{7,7} (\omega _n )|^2, }  \\
\end{array}
\label{eq:E.13}
\end{equation}
\begin{equation}
\begin{array}{*{20}l}
   {\langle a'_1 (\omega _n )a'_2 (\omega _n )\rangle  = } &\!\! {(Q_{\text{th}}  + 1)\!\!\!\!\!\!\sum\limits_{k = 1,3,4,5}\!\!\!\!\!\!\! S_{3,k} (\omega _n )S_{7,k} ^* (\omega _n )}  \\
   {} &\!\! { + (N_{\text{th}}  + 1)\!\!\!\sum\limits_{k = 2,6}\!\! S_{3,k} (\omega _n )S_{7,k} ^* (\omega _n )}  \\
   {} &\!\! { + Q_{\text{th}} S_{3,7} (\omega _n )S_{7,7} ^* (\omega _n ),}  \\
\end{array}
\label{eq:E.14}
\end{equation}
\begin{equation}
\begin{array}{*{20}l}
   {\langle a_1 ^{\prime\dag}  (\omega _n )a_2 ^{\prime\dag}  (\omega _n )\rangle  = } &\!\! {Q_{\text{th}}\!\!\!\!\!\! \sum\limits_{k = 1,3,4,5}\!\!\!\!\!\!\! S_{3,k} ^* (\omega _n )S_{7,k} (\omega _n )}  \\
   {} &\!\! { + N_{\text{th}} \!\!\!\sum\limits_{k = 2,6}\!\! S_{3,k} ^* (\omega _n )S_{7,k} (\omega _n )}  \\
   {} &\!\! { + (Q_{\text{th}}  + 1)S_{3,7} ^* (\omega _n )S_{7,7} (\omega _n ),}  \\
\end{array}
\label{eq:E.15}
\end{equation}
which causes the block matrices of the covariance matrix to simplify to the forms shown in \Eq{30} and \Eq{31}, which give the logarithmic negativity as prescribed in \Sec{II.C}.
\section{\label{sec:App.F}Numerical Methods}
\subsection{\label{sec:App.F.1}Trapezoidal Rule for Scalar Integrals}
We approximated integrals of scalar functions $f(t)$ with the trapezoidal rule,
\begin{equation}
\int_{t_0 }^t {f(t')dt} ' \approx {\textstyle{{\Delta t} \over 2}}\left( {f(t_0 ') + f(t_n ') + 2\sum\limits_{k = 1}^{n - 1} {} f(t_k ')} \right)\!,
\label{eq:F.1}
\end{equation}
with abbreviations
\begin{equation}
\Delta t \equiv {\textstyle{{t - t_0 } \over n}}\;\;\;\text{and}\;\;\;t_k ' \equiv t_0  + k\Delta t,
\label{eq:F.2}
\end{equation}
and $n$ is the number of trapezoidal regions of equal width into which the integral is divided, and is taken as large as it needs to be for a given integral to converge.
\subsection{\label{sec:App.F.2}Suzuki-Trotter Approximation for Propagators}
Here we expand the time-ordered exponentials of form
\begin{equation}
\tau (t,t') \equiv \tau \left\{ {e^{ - i\int_{t'}^t {M(t'')dt''} } } \right\},
\label{eq:F.3}
\end{equation}
where $\tau\{\cdot\}$ is the time-ordering operator.  The approximation formula we used for \Eq{F.3} is, from \cite[]{Suzu},
\begin{equation}
\begin{array}{*{20}l}
   {\tau (t,t')} &\!\! { = \mathop {\lim }\limits_{N \to \infty } \prod\limits^ \leftarrow  {}_{k = 1}^N e^{ - iM(t' + k{\textstyle{{t - t'} \over N}}){\textstyle{{t - t'} \over N}}} }  \\
   {} &\!\! { \approx e^{ - iM(t_N '')\Delta t}  \cdots e^{ - iM(t_2 '')\Delta t} e^{ - iM(t_1 '')\Delta t}, }  \\
\end{array}
\label{eq:F.4}
\end{equation}
where the product's arrow means it grows leftwards, and
\begin{equation}
\Delta t \equiv {\textstyle{{t - t'} \over N}}\;\;\;\text{and}\;\;\;t_k '' \equiv t' + k\Delta t,
\label{eq:F.5}
\end{equation}
where notice that the product starts on $k=1$, not $0$, and $N$ is as large as it needs to be for convergence.
\end{appendix}
%
\end{document}